# Ferroelectric properties controlled by magnetic fields in DyMn$_2$O$_5$

D. Higashiyama[1], N. Kida[3], S. Miyasaka[1], T. Arima[2,3] and Y. Tokura[1,3,4]

[1] *Department of Applied Physics, University of Tokyo, Tokyo 113-8656, Japan*

[2] *Institute of Materials Science, University of Tsukuba, Tsukuba 305-8573, Japan*

[3] *Spin Superstructure Project (SSS), ERATO, Japan Science and Technology Agency (JST), c/o National Institute of Advanced Industrial Science and Technology, Tsukuba, 305-8562, Japan.*

[4] *Correlated Electron Research Center (CERC), National Institute of Advanced Industrial Science and Technology (AIST), Tsukuba, 305-8562, Japan.*

Cross-correlation between magnetic and dielectric properties has been attracting renewed interest because of the fundamental as well as technological importance of controlling the electric (magnetic) polarization by an external magnetic (electric) field. Here, we report the novel phenomenon that an external magnetic field induces and/or modifies ferroelectric states in a magnetic material, DyMn$_2$O$_5$. Measurements of temperature dependence of hysteretic polarization curves, pyroelectric current, specific heat, optical second harmonic generation, and x-ray superlattice peaks have revealed successive phase transitions between 43 K and 4 K, accompanying three ferroelectric phases. The zero-field lowest-temperature phase (<8 K) induced by the Dy-moment ordering is a reentrant paraelectric state, but is turned to a ferroelectric state with increasing the magnetic field. The phenomenon is closely related to the metamagnetic transitions of the Dy *f*-moment, indicating that all the ferroelectric phases of this material are strongly tied to the antiferromagnetic Mn spin structure affected by the *f-d* exchange interaction. The electric phase diagram for DyMn$_2$O$_5$ is presented in the plane of temperature and magnetic field.

# I. INTRODUCTION

Controlling the electric polarization by a magnetic field, or inversely the magnetization by an electric field has long been an important and interesting topic, termed magneto-electric (ME) effect. The possible effect was first predicted by Pier Curie[1] in 1894, but its reality in a substantial material $Cr_2O_3$ was pointed out by Dzyaloshinski[2] in 1959 and experimentally demonstrated by Astrov[3] in 1960. Since then, the ME effect has been confirmed on many materials[4-6]. Studies of the ME effect have been providing fundamentally important information about the cross-correlation between the magnetic and electric polarizations, yet its magnitude has remained too small for actual technological application of this intriguing phenomenon to be realized. A key to the breakthrough is believed to be the use of *multiferroics*, where the ferroic orders of (anti)ferromagnetism and (anti)ferroelectricity coexist.

Although the both ferromagnetic and ferroelectric materials are quite rare, there have been reported several families of magnetic ferroelectrics in which the spontaneous electric-polarization can be controlled to a considerable extent by an external magnetic field. The first example is a weakly-ferromagnetic boracite $Ni_3B_7O_{13}I$ as reported by Schmid and coworkers[7], in which the rotation of the magnetic field by 90º causes a 180º switching of the spontaneous electric polarization. Quite recently, Kimura *et al.*[8] have demonstrated a new approach to the magnetic field control of the spontaneous polarization. In perovskite type (orthorhombically distorted) $TbMnO_3$, the ferroelectric polarization emerges upon the incommensurate (IC) to commensurate (C) transition of the antiferromagnetic spin order. At a lower temperature than 10 K, the Tb 4*f* moments undergo the ordering due to the coupling with the Mn d-electron spins. Then, a magnetic field induces the metamagnetic transition related to the flop of the Tb moments, perhaps alters the commensurate spin structure of the Mn spins, and resultantly changes the 90º flop of the spontaneous polarization. The key ingredients for the success of the magnetic field control are; (1) the commensurate *d*-electron spin structure that couples magneto-elastically with the lattice modulation producing the spontaneous polarization, and (2) the *f-d* exchange interaction that amplifies the action of the external magnetic field to alter the *d*-electron spin state. In this context, a similar way of magnetic-field control may be applied to other rare-earth transition-metal oxides that show magnetic-order induced ferroelectricity. Apart from the family of perovskite $R$MnO$_3$ ($R$= Gd, Tb, Dy)[9], one of the promising candidates for this purpose is a series of $R$Mn$_2$O$_5$ (here $R$ being rare-earth ions, Y, or Bi) with magnetically induced ferroelectricity[10]. Here, we report the variation of the ferroelectric phases with application of an external magnetic field in DyMn$_2$O$_5$. The most striking feature found in this study is that the ferroelectric states are successively induced in the paraelectric ground state by applying the magnetic field. This may be viewed as a new type of gigantic ME effect. We present here the electric phase diagram in the plane of temperature and magnetic field for DyMn$_2$O$_5$.

A family of $R$Mn$_2$O$_5$ is known as the magnetic ferroelectrics which undergo the ferroelectric transition at 25-39 K below the antiferromagnetic transition temperature $T_N$=39-45 K[10]. According to the structural study[11,12], the orthorhombic crystal (*Pbam*) at room temperature is composed of Mn$^{4+}$O$_6$ octahedron and Mn$^{3+}$O$_5$ pyramid units, as shown in Fig. 1. The octahedra form the chains along the *c*-axis, while sharing their edges. On the other hand, a pair of the pyramids links these chains within the *ab* plane. The emergence of ferroelectricity is believed to arise from the pyramidal Mn$^{3+}$ sites[13]. The magnetic structure of DyMn$_2$O$_5$ was determined in terms of neutron diffraction by Wilkinson *et al.*[14]. At 4.2 K the antiferromagnetic state is characterized by the two magnetic propagation vectors (1/2 0 0) and ((1+ )/2 0 1/4) with ~0.001. The Mn$^{4+}$ octahedra form ferromagnetic chains with a spin moment of 2.0 $\mu_B$, while the Mn$^{4+}$ spins, oriented at a right angle to the ferromagnetic component, are also antiferromagnetically modulated with amplitude of 2.6 $\mu_B$ and a periodicity of 4*c*. With increasing temperature, it was observed[14] that the (*h*/2 *k* 0) magnetic superlattice peaks (and thus the ferromagnetic component in each Mn$^{4+}$ chain) disappear at 8 K (perhaps corresponding to $T_X$ that we define later), which should result from disordering of the Dy moments. The magnetic scattering corresponding to the ((1+ )/2 0 1/4) modulation persists above 8 K. The ((*h*± )/2 0 *l*±1/4) split peaks merge to (*h*/2 0 *l*±1/4) at 18 K(=$T_3$), and finally these antiferromagnetic peaks disappear about 44 ± 2 K(=$T_N$). DyMn$_2$O$_5$ undergoes such complicated successive magnetic transitions with varying temperature, indicating that there are competing exchange interactions among the Mn spins as well as coupling between the Mn spins and Dy moments. Here, we fully resolve the interrelation of the ferroelectricity/paraelectricity with the respective spin-ordered phases by measurements of dielectric constant, spontaneous polarization, specific heat, and magnetic susceptibility.

## II. EXPERIMENT

Single crystals of $DyMn_2O_5$ (several mm in size) were grown with use of $PbO\text{-}PbF_2$ flux[15]. The oriented samples of rectangular shape with well-developed crystal surfaces of (001), (010), and (100) were prepared with use of back-Laue photographs. For the measurement of *P*-*E* curves, the triangular wave electric field (*E*) of 10 Hz was applied on the sample capacitor. The dielectric constant was measured by a commercial LCR meter (HP4274A) operated at 1 kHz on cooling the sample at a rate of 5 K/min. The specific heat was measured by a relaxation method in the cooling run at a rate of ~0.3 K/min. Concerning the derivation of the spontaneous polarization ($P_s$) from the pyroelectric measurement, special caution should be paid to the present material. This is because the lowest-temperature phase of $DyMn_2O_5$ is the zero-$P_s$ state (see III. Results and discussion) and accordingly the conventional procedure of the electric-field cooling and the subsequent zero-electric-field warming procedure for the pyroelectric current measurement is invalidated. Therefore, we have measured the pyroelectric current with use of an electrometer (KEITHLEY6517A) while applying the relatively small electric field (200 kV/m) on the cooling run. Also when the magnetic field was scanned at a rate of 38 Oe/s, the pyroelectric current was measured while applying the electric field. Therefore, the polarization value for the final state in the course of the temperature or magnetic-field scan may be slightly affected by a tiny leak current and give some uncertainty in the obtained value. Nevertheless, such an electric-field-applied measurement is indispensable for the case where the zero-polarization state is involved during the temperature and/or magnetic field scans as in $DyMn_2O_5$.

We have also adopted the second harmonic (SH) generation technique to confirm the breaking of centrosymmetry induced by the ferroelectric order. The s-polarized fundamental light (the electric field *E* being set to parallel to the ferroelectric *b*-axis) of 1 mW from a Ti:sapphire regenerative amplifier system (the center wavelength of 800 nm, the pulse width of 130 fs, and the repetition rate of 1 kHz) was irradiated on the ab surface with the focused spot diameter of 0.5 mm and the incidence angle of 40º. The generated s-polarized (//*b*) SH light was detected by a photomultiplier tube attached to a monochromator and a boxcar averager. The sample was once cooled to 20 K at a rate of 2 K/min in a poling field of 1.5 MV/m. After the poling field is removed at 20 K, SH measurements were performed in the warming and cooling runs.

The superlattice structure associated with the ferroelectric order in the spin ordered state were searched for in the reciprocal space of ($h$ 0 $l$) in terms of synchrotron-radiation X-ray diffraction. The measurements were performed by using a Huber six-axis diffractometer at Beam Line-4C, Photon Factory of KEK, Japan. The incident X-ray was monochromated at 13 keV by utilizing a Si (1 1 1) double-crystal monochromator and focused on a (001) surface of a single crystal mounted in a closed-cycle He refrigerator. To reduce the contamination of luminous X-ray, a pyrographite (0 0 2) monochromator was installed in front of the scintillation counter.

## III. RESULTS AND DISCUSSION

### A. Ferroelectric and paraelectric states coupled with *d*-spin and *f*-moment orders

Let us first present the unambiguous evidence for the ferroelectricity of $DyMn_2O_5$ crystal at low temperatures. Figure 2 exemplifies the *P-E* hysteretic curves at some selected temperatures in the cooling process. The electric field was applied along the *b*-axis, since the spontaneous polarization ($P_s$), if any, is always observed along the *b*-axis. Above 40 K, even below the magnetic ordering temperature ($T_N$=43 K), the *P-E* curve is characteristic of paraelectrics, showing no $P_s$. As decreasing temperature below 40 K, the clear hysteresis loop characteristic of ferroelectrics is observed, and the remanent polarization as well as the coercive force is increased to reach maximum around 26 K. The value of the *P* at zero *E* is plotted against temperature with closed circles in Fig. 3(d) to show the temperature-variation of $P_s$. Below 13 K, however, the coercive force appears to steeply increase, and accordingly the *P* does not show the saturation even at 1.4 MV/m and the *P-E* curve shows the rounded shape as a whole. The obtained low $P_s$ value at zero *E* (see also Fig. 3(d)) in this temperature range thus underestimates the real $P_s$ that should be much larger, as discussed later in comparison with other $P_s$ data derived by the pyroelectric current (Fig. 3(e)). As a remarkable fact, the *P-E* curve below 7 K again shows no ferroelectric *P-E* curve. This indicates that the lowest-temperature phase below 7 K is a reentrant paraelectric state with no $P_s$. The *antiferroelectric* state might be possible, but the field-induced polarization cannot be observed up to 1.8 MV/m, above which the crystal shows dielectric breakdown. In accord with the emergence of the ferroelectric state, we have confirmed the second-harmonic (SH) generation of the b-polarized incident light. The SH light intensity, that is expected to be in proportion with square of $P_s$, was observed to decrease to the background level in the paraelectric phases at the temperatures below 7 K (down to 4.2 K) as well as above 40 K, as shown with open circles in Fig. 3(d). Here, we assumed the temperature-independent background signal above 40 K, perhaps arising from surface effects, as the zero level of the ferroelectricity-induced SH signal.

Figure 3 compares the temperature dependencies of (a) the x-ray superlattice peak observed at (0 0 4.5), (b) the specific heat $C$ divided by temperature $T$, (c) the dielectric constant measured with applied electric-field (1 kHz) along the respective crystallographic axes, (d) $P_s$ as deduced from the *P-E* curves (Fig. 2), and (e) $P_s$ as deduced by the pyroelectric current during the electric-field (200 kV/m) cooling. (Note the change of the temperature scale on the abscissa below and above 20 K.) From these comprehensive measurements, we could reveal the five successive phase transitions occurring in $DyMn_2O_5$. First, the compound undergoes the antiferromagnetic transition, perhaps associated with the IC spin order, at $T_N$=43 K, and then the ferroelectric transition at $T_1$=40 K immediately below the $T_N$. These critical temperatures are close to each other but definitely distinguished as the double peaks in the specific heat data shown in Fig. 3(b). Around $T_1$, only the *b*-axis component of the dielectric constant shows a sharp anomaly as demonstrated in Fig. 3(c), in accord with the presence of the spontaneous polarization along the *b*-axis in the lower-lying phases. The x-ray superlattice reflection at (0 0 4.5) is observed to increase in lowering temperature from $T_1$, but tends to decrease in intensity below 28 K (=$T_2$) and then finally disappears around 13 K (=$T_3$), as shown in Fig. 3(a) and Fig. 4. The emergence of such a C lattice modulation (see the Fig. 4 for the observed peak profiles) is likely to correspond to the magnetic modulation with the propagation vector (1/2 0 1/4) reported in the temperature range between 18 K and 44 K by Wilkinson *et al.*[14]. As observed in the case of the ferroelectric perovskite-type $R$MnO$_3$ [8], the strong exchange-striction effect inherent to the Mn-O network produces the lattice modulation with twice of the modulation vector of the spin modulation. In analogy to the improper ferroelectrics like Rb$_2$ZnCl$_4$ and related compounds[16,17] as well as to perovskite $R$MnO$_3$, the IC to C transition of lattice modulation may produce the parasitic ferroelectricity also in this compound. What is unique in this system is the presence of multiple ferroelectric (FE1, FE2, and FE3) phases perhaps associated with the reordering of the spin state. All these ferroelectric phases show the spontaneous polarization only along the *b*-axis.

The FE1-to-FE2 and FE2-to-FE3 transitions at $T_2 \sim 28$ K and $T_3 \sim 13$ K (on cooling), respectively, are clearly discerned as the sudden changes of $\boldsymbol{P}_s$ in Figs. 3(d) and 3(e). These transition temperatures are to some extent, up to ± 2 K, sample-dependent, possibly due to slight oxygen offstoichiometry. The former transition is also manifested as the clear anomaly of the specific heat shown in Fig. 3(b). A tiny discontinuity of the specific heat around 16 K (indicated by an open triangle in Fig. 3(b)) may correspond to the FE2-to-FE3 transition. The phase transition between FE2 and FE3 shows a thermal hysteresis (a difference in the transition points on cooling and warming, $\Delta T_3 \sim 4$ K), indicating its first-order nature. The FE3 phase shows the large coercive force (>0.6 MV/m) as shown in Fig. 2, but the $\boldsymbol{P}_s$ estimated from the electric-field cooling pyroelectric current shows the comparable value in other FE phases (see Fig. 3(e)). The respective ferroelectric phases and their variation appear to be closely related to the successive emergence of the spin order with different propagation vectors ($k_s$), and the part of the lattice modulation caused by the commensurate ($k_s$=(1/2 0 1/4)) spin order can be seen as the superlattice peaks at $2k_s$ in the FE1 and FE2 phases. However, we could not observe ( 0 9/2) superlattice in the FE3 phase (see the 11.8 K curve in Fig. 4(a)), which implies that the displacement of Mn and O may change in direction and become confined in the *ab* plane. From the present combined measurements, however, the spin state transitions seem to be even more complex than reported in the literature[14], which may miss the spin reorientation transitions corresponding to the PE($<T_N$)-FE1 and FE1-FE2 transitions at $T_1$ and $T_2$, respectively, and need to be carefully reexamined to correlate with the five ferroelectric/paraelectric phases.

With further lowering temperature, the compound undergoes the phase transition from the FE3 to the new paraelectric state, here referred to as the phase X, around $T_X$=7 K. This transition shows the thermal hysteresis ($\Delta T_X \sim 1$ K). The first-order nature of the transition is also evidenced by a steep drop of the $\boldsymbol{P}_s$. The seemingly residual $\boldsymbol{P}_s^{(pyro)}$ in the phase X as seen in Fig. 3(e), as contrasted by zero $\boldsymbol{P}_s^{(P-E)}$ in Fig. 3(d), might be due to the subsisting coexistence of the FE3 state due to the first-order nature of the transition, but more plausibly due to the leak current arising from the electric-field cooling measurement of the pyroelectric current. The conventional electric-field-cooling and zero-electric-field-warming procedure cannot be applied to the present material, since the lowest-temperature phase X shows no $P_s$ (see II. Experiment for details). This paraelectric phase X is obviously related to the ordering of the Dy $f$-electron moments (and resultant modification of $d$-electron spins) as manifested by the large peak in the specific heat (Fig. 3(b)). The ordering of the Dy moments also affects the spin ordering of $Mn^{3+}$ and $Mn^{4+}$, as revealed by Wilkinson et al.[14]. Upon the $f$-moment ordering, the spin structure in a $Mn^{4+}$ chain along the $c$-axis changes from the purely antiferromagnetic state with a periodicity of $4c$ to the ferrimagnetic one. The exchange striction and parasitic ferroelectricity can result only from the antiferromagnetic component. This implies that the application of an external magnetic field causing the flop transition of the $f$-moments may cause the change in the $d$-spin structure and accordingly the electric states.

### B. Magnetic-field induced phase transitions

The magnetic field effect on the temperature dependence of $P_s$ (//$b$) is shown in the Fig. 5(b). The $P_s$ value was deduced again by the measurement of the pyroelectric current in cooling the crystal while applying the electric field (200 kV/m) along the $b$-axis in a constant magnetic field (0-3 T) parallel to the $a$-axis. (The magnetic easy-axis appears to lie within the $ab$ plane.) The critical temperature $T_X$ for the transition between FE3 and X is shifted to lower temperature with the magnetic field, and above 1.5 T the phase X is extinguished and replaced by the FE3 phase. Furthermore, the transition temperature $T_3$ between the FE3 and FE2 also shows a shift toward low temperature with the applied magnetic field, and above 2.5 T the FE3 phase also disappears and is transformed into the FE2 phase. Thus, we can present the electric phase diagram in the plane of temperature and magnetic-field, as shown in Fig. 5(a). A large field-shift is observed for the lower-lying transition temperatures, $T_X$ and $T_3$, whereas very little for $T_2$ and $T_1$. This implies that both the FE3 and the nonpolar phase X are affected by the coupling between the $d$-spins and $f$-moments. Namely, these ferroelectric/paraelectric states at low temperatures are under the influence of $f$-moment arrangement that is amenable to the external magnetic field.

The magnetic-field induced phase transition as the result of strong coupling between magnetism and ferroelectricity is clearly manifested also in the isothermal curves of the magnetic-field induced electric-polarization. Figure 6 exemplifies the magnetic field (***H***//*a*) scan at 3 K. Figure 6(b) shows the polarization change derived by the integration of the pyroelectric current while the electric field (200 kV/m) was always applied along the *b*-axis. In the case of zero applied electric-field, the direction of the induced polarization was observed not to be uniquely nor controllably fixed, which should arise from zero spontaneous polarization in the starting phase X. The magnetic-field induced polarization appears in a two-step manner as indicated by vertical dashed lines. These transitions correspond to the transitions to FE3 and FE2, respectively, as noticed from the phase diagram shown in Fig. 5(a). In the ***H*** decreasing run, the two step-anomalies are also observed but respectively accompany the large hysteresis in the transition magnetic-field. After the completion of the ***H*** increasing and decreasing run, the total displacement (pyroelectric) current cannot completely cancel out, producing some seemingly finite ***P*** value at the zero magnetic-field. This might partly be due to some residual FE3 phase due to the strong first-order nature of the transition, and/or due to some leak current because of the procedure of the electric-field application. It is to be noted here that the sign of the ***P*** in the magnetic-field induced ferroelectric states remains unchanged with the reversal of ***H*** direction, as far as the small electric field is applied along the same direction during the ***H*** scans. Corresponding to these transition magnetic fields, the magnetization (***M***-***H***) curves (Fig. 6(a)) show the metamagnetic anomalies corresponding to the flop of the Dy *f*-moments. The first one associated with the clear hysteresis loop corresponds to the X-FE3 transition. The anomaly corresponding to FE3-FE2 transition is not so clear in the bare ***M***-***H*** curve but discernible in its derivative (d***M***/d***H***-***H***) curve. Those hysteretic behaviors are also in accord with those of the ***P***-***H*** curves shown in Fig. 6(b), as guided by the vertical dashed lines. This confirms again that the induced electric polarization arises from the phase transitions mediated by the Zeeman energy but not from the conventional linear ME effect in which the sign reversal should occur upon the reversal of the ***H*** direction.

## IV. SUMMARY

The three distinct ferroelectric phases (FE1,2,3) and the lowest-lying reentrant paraelectric phase (X) have been identified in $DyMn_2O_5$. All the ferroelectric phases emerge in the spin ordered phase perhaps with the commensurate propagation vectors. The coupling between the magnetism and the lattice distortion as the source of the ferroelectricity is also evidenced by the presence of the superlattice peak at twice of the magnetic propagation vector in FE1 and FE2 phases. Furthermore, the lower-lying FE3 and paraelectric X phases can be altered by application of external magnetic fields, indicating that these low-temperature dielectric properties are affected also by the Dy *f*-moment order coupled with the Mn *d*-spin. In particular, the paraelectric ground state (X) shows the successive ferroelectric transitions to FE3 and then to FE2 with increasing the magnetic field. The magnetic-field induced ferroelectric polarization of about 1 $mC/m^2$ is viewed as the gigantic magneto-electric effect. The similar magnetic-field control of the dielectric phase should be ubiquitous in various rare-earth (*R*) manganese oxides including families of *R*$Mn_2O_5$ and *R*$MnO_3$, in which the Mn *d*-electron spins with the large exchange-striction effect as the source of ferroelectricity couple with the magnetic-field controllable *f*-moments of *R* ions.

## ACKNOWLEDGMENTS

The authors would like to thank A. Sawa and S. Horiuchi for their help in experiment, and K. Kohn and T. Kimura for fruitful discussions. This work was partly supported by Grant-In-Aids for Scientific Research from the MEXT, Japan.

*Figure legends*

Fig. 1. Crystal structure of DyMn$_2$O$_5$ : A view along the *c*-axis (upper panel) and along the *a*-axis for the sub-cells labeled 1,2,3 and 4 (lower panel).

Fig. 2. The polarization (***P***) vs. electric-field (***E***, 10Hz) curves at various temperatures in the spin ordered state of DyMn$_2$O$_5$. The ferroelectric states show the hysteretic ***P-E*** curves (at 33.0, 16.6, and 13.0 K in the figure), while the linear ones stand for the paraelectric states (at 40.0 and 4.4 K).

Fig. 3. The temperature dependence of several quantities related to the magnetic and/or dielectric phase transitions in DyMn$_2$O$_5$; (a) the integrated intensity (*I*) of x-ray superlattice peak at (0 0 4.5) shown in Fig. 4, (b) the specific heat (*C*) divided by temperature (*T*), (c) the dielectric constants for applied electric fields (1 kHz) along the *a*, *b* and *c*-axis, (d) the spontaneous polarization (***P***$_s^{(PE)}$, closed circles) along the *b*-axis obtained from the polarization vs. electric-field (***P-E***) curves shown in Fig. 2, and (e) the polarization (***P***$_s^{(pyro)}$) along the *b*-axis obtained by the measurement of pyroelectric current under an applied electric field of 200 kV/m. All the above temperature scans were done in the cooling run. In the panel (d), the temperature dependence of the second-harmonic light intensity $I_{SHG}$ is shown with open circles. (See the text for the experimental details.)

Fig. 4. X-ray superlattice peak profiles in (a) (*h* 0 4.5) and (b) (0 0 *l*) scans at various temperatures in DyMn$_2$O$_5$.

Fig. 5. (a) The electric phase diagrams (X and PE: paraelectric; FE1, FE2, FE3: ferroelectric) in the plane of temperature and magnetic field as obtained from the measurement on (b) the temperature profile of the polarization (//*b*) at various magnetic fields. The polarization was deduced by the integration of the pyroelectric current in the electric-field (200 kV/m) cooling run. The electric polarization and the applied magnetic

field are parallel to the *b*- and *a*-axis, respectively. Open triangle with horizontal bars in (a) means that the transition does not appear above 3 K.

Fig. 6. (a) Magnetization (*M-H*) curves and (b) magnetoelectric effect (*P-H* curves) in DyMn$_2$O$_5$ at 3 K. The applied magnetic field *H* was parallel to the *a*-axis. The magnetization curve shows a clear hysteresis loop around 1.5 T, corresponding to the metamagnetic transition of the Dy *f*-moment. The two metamagnetic anomalies are discerned in the *H*-derivative of magnetization curves as shown by vertical dashed lines in (a) on the *H*-increasing and -decreasing runs, respectively. (The sign of d*M*/d*H* in an *H*-decreasing run is inverted.) The magnetic field dependence of the electric polarization shown in (b) was measured in an applied electric field of 200 kV/m. In accord with the magnetic transitions, the electric polarization shows the step-like increases (decreases) with increasing (deceasing) *H*. A large *H*-hysteretic behavior is also seen in the *P-H* curves (b), corresponding to that of the *M-H* curve shown in (a), as guided by the vertical dashed lines.

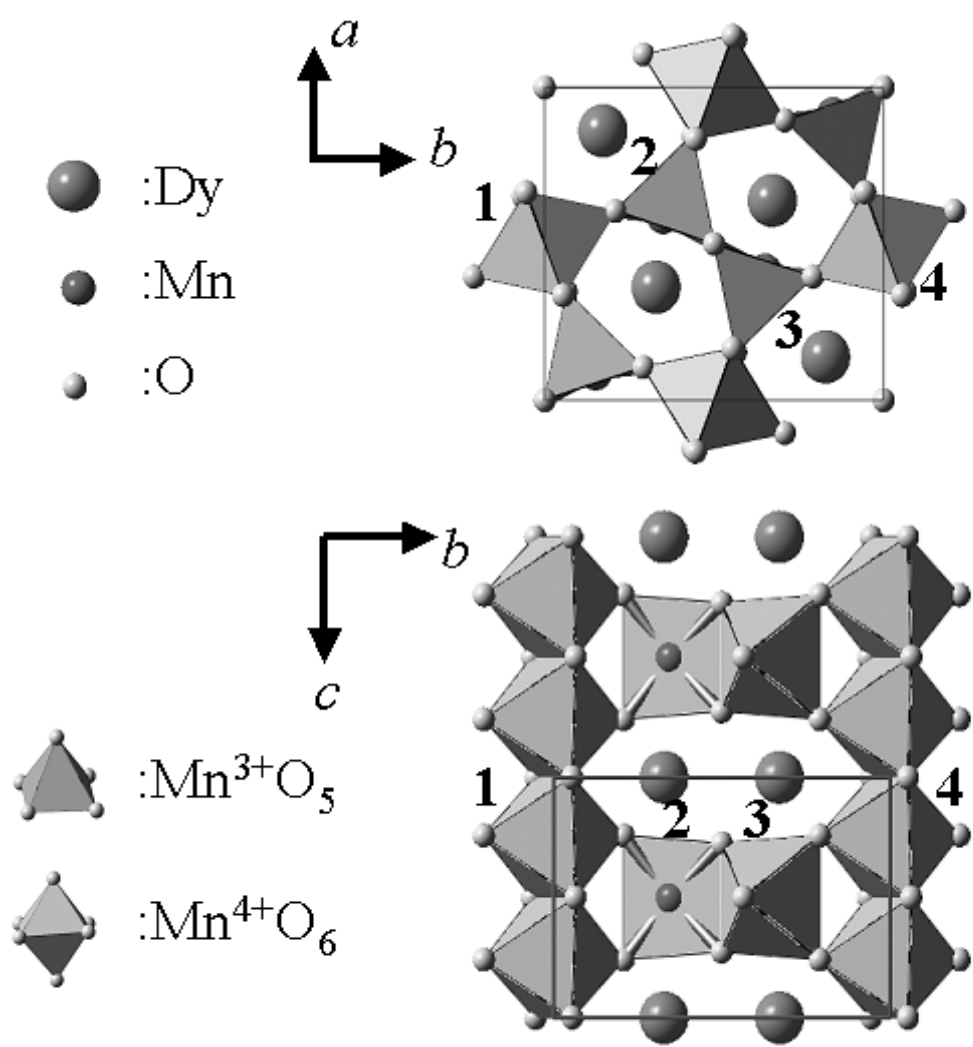

Fig. 1    D. Higashiyama, et al.

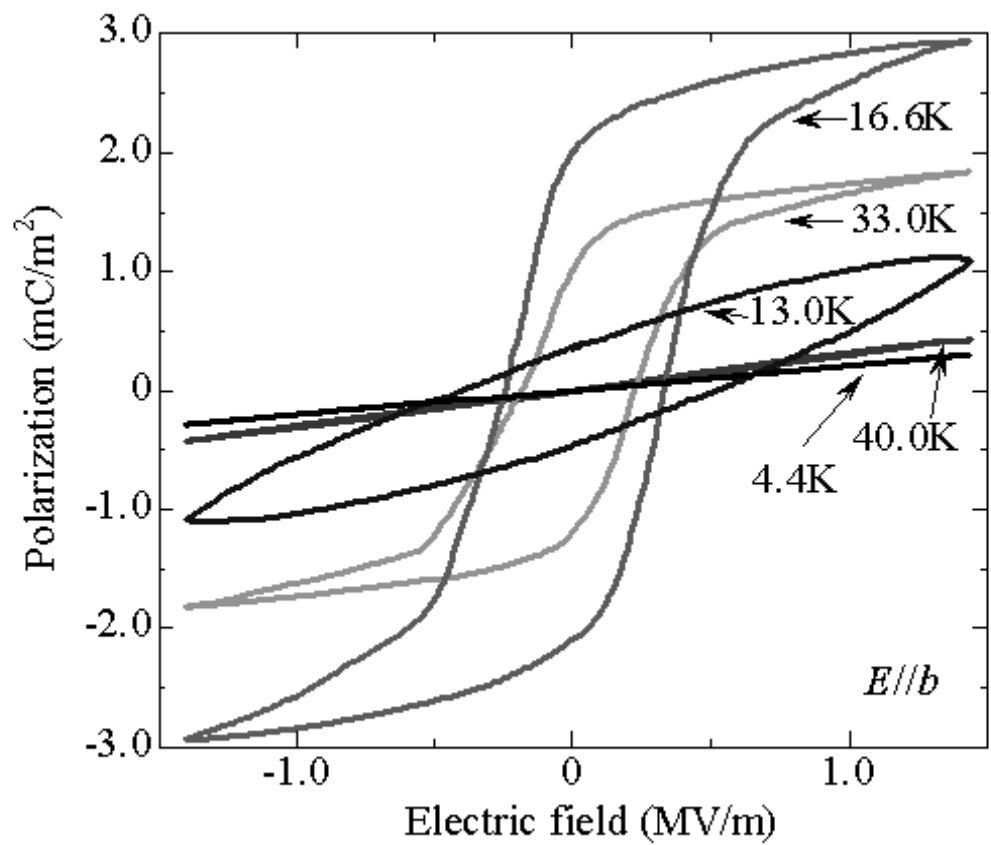

Fig. 2  D. Higashiyama, et al.

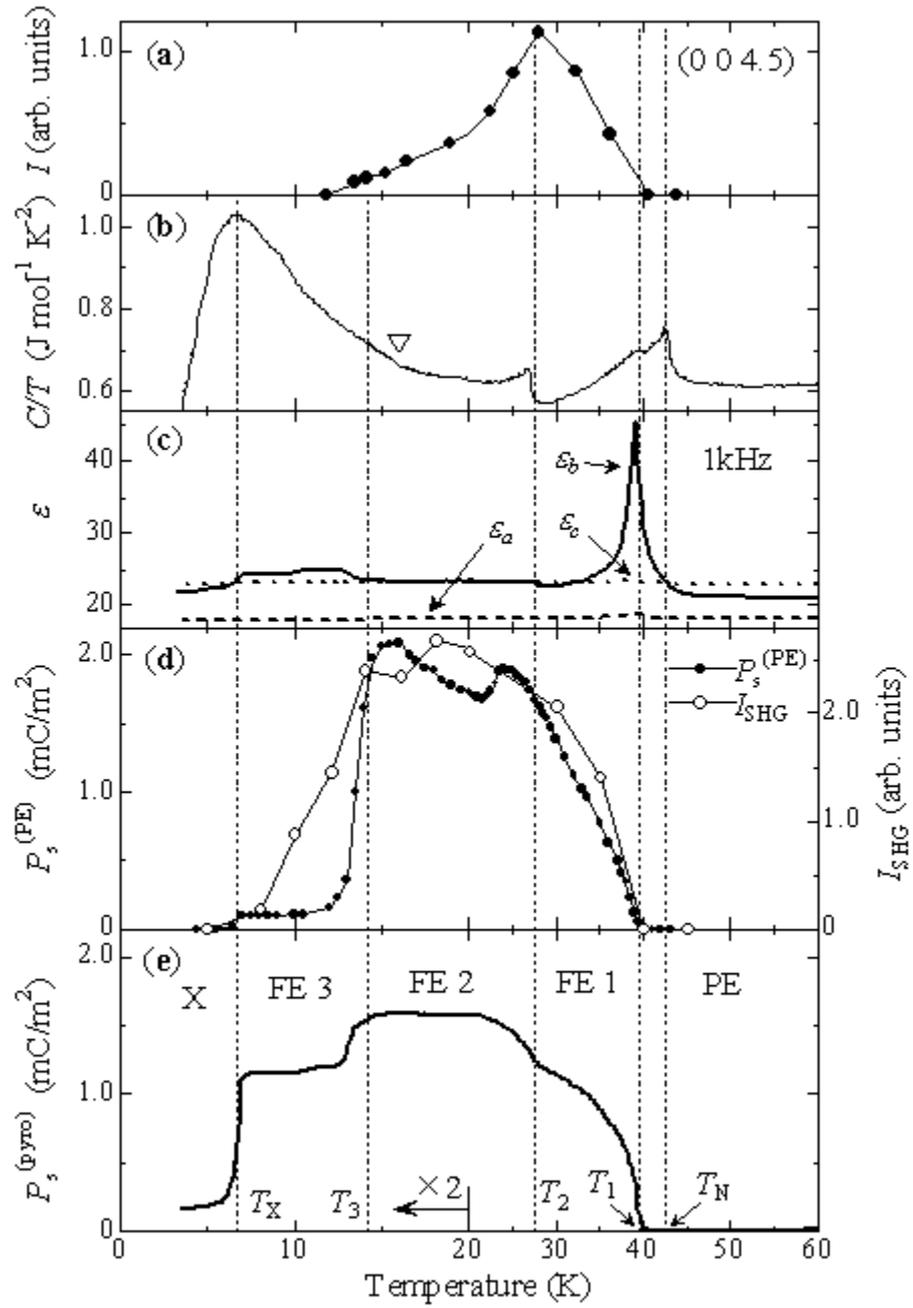

Fig. 3  D. Higashiyama, et al.

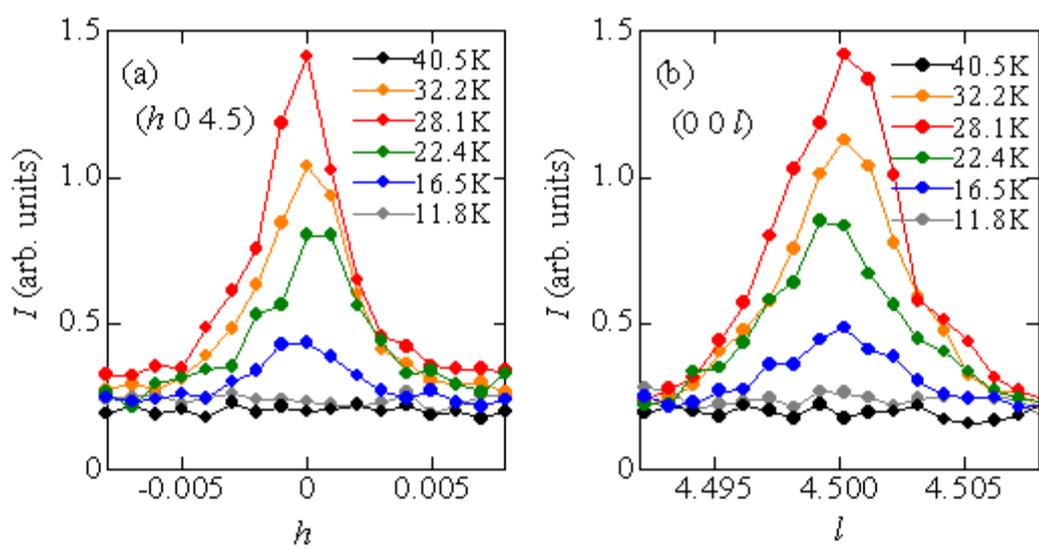

Fig. 4  D. Higashiyama, et al.

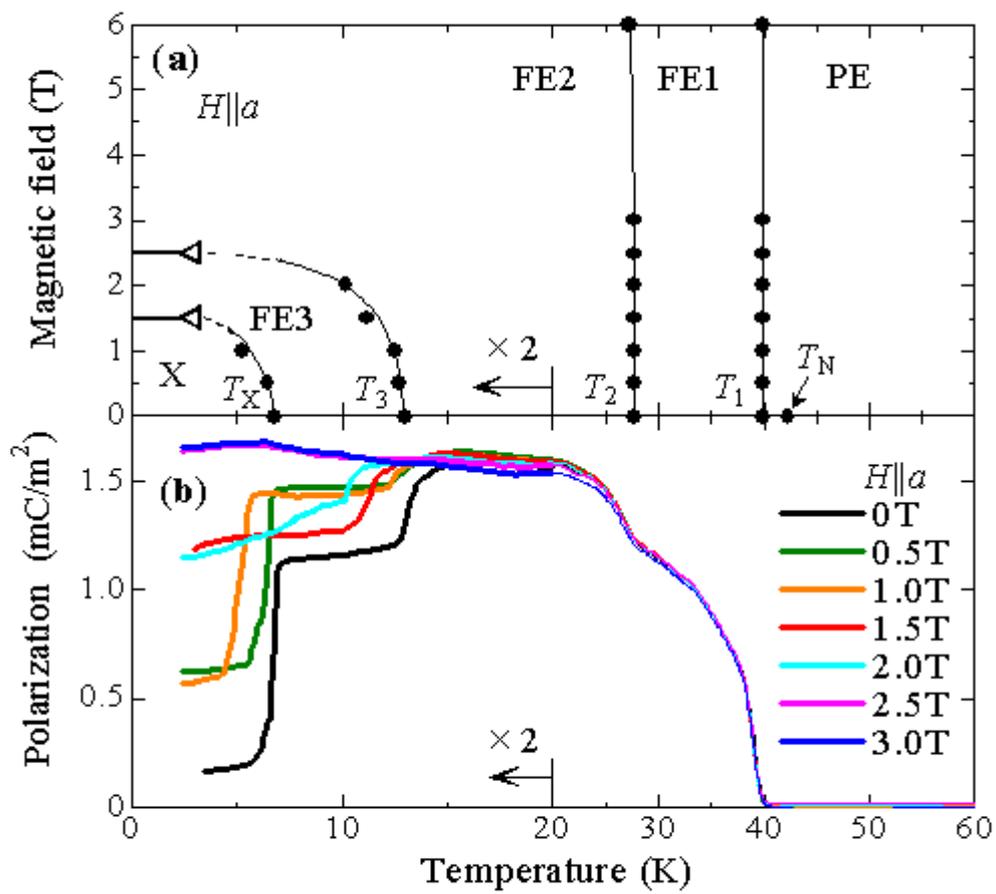

Fig. 5    D. Higashiyama, et al.

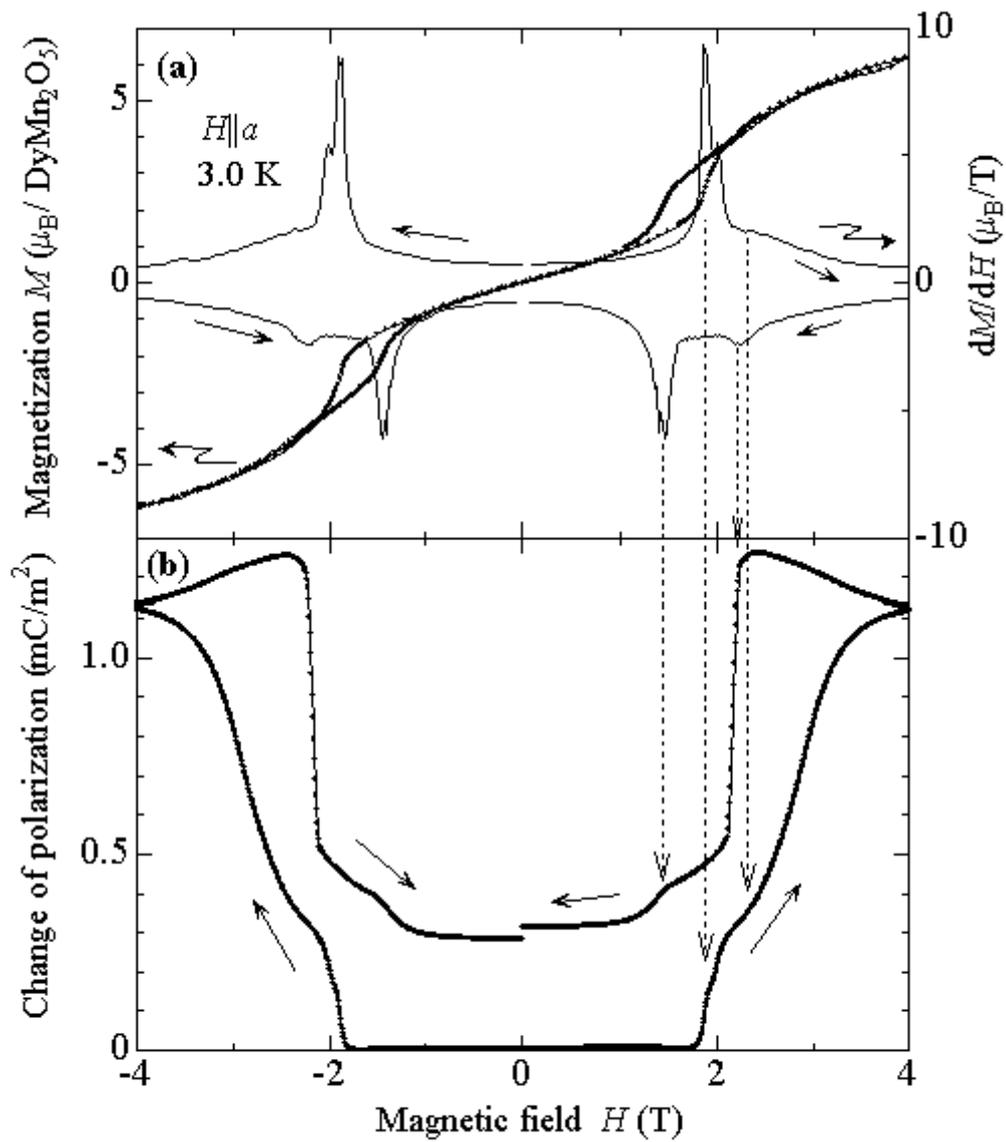

Fig. 6    D. Higashiyama, et al.